\documentclass{article}
\usepackage{epsfig}
\usepackage{amssymb}

\begin{document}
\begin{large}
\pagenumbering{arabic}

\rightline{February 23, 2003}
%\rightline{(4th version)}

\vskip 1.0cm

\begin{center}
\bigskip
{\LARGE{\bf Study of the colour singlet model with $k_T$-factorization\\
  in inclusive $J/\psi$ production at CERN LEP2}}\\

\vspace{1.5cm}

{\large 
A.V. Lipatov\footnote{E-mail address: lipatov@theory.sinp.msu.ru}\,

{\it Physical Department,\\
 M.V. Lomonosov Moscow State University,\\
119992 Moscow, Russia\/}\\[3mm]}

\vskip 0.5cm

N.P. Zotov\footnote{E-mail address: zotov@theory.sinp.msu.ru}

{\it D.V.~Skobeltsyn Institute of Nuclear Physics,\\ 
M.V. Lomonosov Moscow State University,\\
119992 Moscow, Russia\/}\\[3mm]

\end{center}

\vspace{1.0cm}

\begin{center}
{\bf{Abstract}}
\end{center}
\bigskip

We calculate the cross section and the transverse momentum distribution
of inclusive $J/\psi$ production in $\gamma \gamma$ 
collisions at CERN LEP2 within the colour singlet model and the 
$k_T$-factorization approach, including both direct and resolved photon 
contributions. The unintegrated gluon distribution in the photon
is determined, using the Bl\"umlein's prescription for unintegrated 
gluon distribution in a proton. We compare our 
theoretical predictions with preliminary data taken by the DELPHI
 collaboration at LEP2. In addition, we present our predictions for the 
$J/\psi$ polarization properties.

\vspace{1.0cm}
\section{Introduction} \indent 

Heavy quark and quarkonium production at high energies has been
vigorously studied in recent years from both theoretical and
experimental viewpoint. At the modern collider conditions 
these processes are so called semihard ones~[1--4]. In such processes 
by definition the hard scattering scale $\mu \sim m_Q$ is large compare 
to the $\Lambda_{{\rm QCD}}$ parameter but on the other hand $\mu$ is much 
less than the total center-of-mass energy: 
$\Lambda_{{\rm QCD}}\ll \mu\ll\sqrt s$. The last condition implies 
that the processes occur in small $x$ region: $x\simeq m_Q/\sqrt s\ll 1$, 
and that the cross sections of heavy quark and quarkonium production 
are determined by the behavior of gluon distributions in the small $x$ 
region.

It is also known that in the small $x$ region it becomes necessary to 
take into account the dependences of the subprocess cross section 
and gluon distribution on a gluon transverse 
momentum $k_T$~[1--4]. Therefore calculations of heavy quark and 
quarkonium production cross sections at HERA, LEP/LHC and other 
collider conditions are necessary to carry out in the so called 
$k_T$-factorization~[2, 3] (or semihard~[1, 4]) QCD approach, which is more preferable 
for the small $x$ region than standard parton model.

The $k_T$-factorization approach is based on Balitsky, Fadin, Kuraev, 
Lipatov (BFKL)~[5] evolution equations. The resummation of 
the terms $\alpha_{S}^n\,\ln^n(\mu^2/\Lambda_{{\rm QCD}}^2)$, 
$\alpha_{S}^n\,\ln^n(\mu^2/\Lambda_{{\rm QCD}}^2)\,\ln^n(1/x)$ and 
$\alpha_{S}^n\,\ln^n(1/x)$ in the $k_T$-factorization approach leads 
to the unintegrated (dependent from $q_T$) gluon distribution 
$\Phi(x,q_T^2,\mu^2)$ which determine the probability to find a 
gluon carrying the longitudinal momentum fraction $x$ and transverse 
momentum $q_T$ at probing scale $\mu^2$. In contrast with the usual 
parton model, to calculate the cross section of a physical process 
the unintegrated gluon distributions have to be convoluted with off 
mass shell matrix elements corresponding to the relevant partonic 
subprocesses~[1--4].

Nowadays, the significance of the $k_T$-factorization approach 
becomes more and more commonly recognized. Its applications
to a variety of photo-, lepto- and hadroproduction processes
are widely discussed in the literature (see~[6] and references cited 
therein).

It is known that $J/\psi$ production at high energies is an 
intriguing subject in modern physics. It is traces back to the
early 1990s, when the CDF data on the $J/\psi$ and $\Upsilon$
hadroproduction cross section revealed a more than an order of magnitude
discrepancy with theoretical expectations. This fact has induced
extensive theoretical investigations. In particular, it was required to 
introduce new additional production mechanism, the 
so-called colour octet (CO) model~[7], where $c\bar c$-pair is 
produced in the color octet state and transforms into final 
colour singlet (CS) state by help soft gluon radiation. Since then,
the color octet model has been believed to give the most likely
explanation of the quarkonium production phenomena, although there are
also some indications that it is not working well. 

One of the problems is connected to the $J/\psi$ photo- and leptoproduction 
data at HERA. For example, the contributions from the color octet 
mechanism to the $J/\psi$ photoproduction contradict the H1 and ZEUS 
data for $z$ distribution~[8--11]. Also in leptoproduction case
the shapes of the $Q^2$, $p_T^2$ and $y^*$ spectra are not reproduced 
by calculations~[12] within the color octet model. The $z$ distributions
calculated later in~[13] contradict the H1 experimental data, too.

Another difficulty of this model refers to the $J/\psi$ polarization 
properties in $p\bar p$-collisions at Tevatron. If, as expected, the dominant contribution comes from the gluon
fragmentation into an octet $c\bar c$ pair, the $J/\psi$ mesons
must have strong transverse polarization~[14--18]. This is in
disagreement with the experimental data~[19], which point to unpolarized or 
even longitudinally polarized $J/\psi$ mesons.

Taking into account the above mentioned problems of color octet mechanism, 
we have considered the inelastic $J/\psi$ meson photo- and leptoproduction at HERA 
within the colour singlet model with $k_T$-factorization~[20]. 
Our theoretical results agree well with H1 and ZEUS experimental data 
for all distributions without any additional $J/\psi$ production mechanisms (such as given by 
the CO model)(see also [21]). Also it was shown that the $k_T$-factorization 
approach gives a correct description of the $J/\psi$ polarization properties in
$ep$ and $p\bar p$ interactions~[20, 22] at the large transverse momenta $p_T$.

Based on these theoretical results, in this paper we consider the inclusive 
$J/\psi$ production at CERN LEP2 conditions within the colour singlet model 
with $k_T$-factorization. There are several motivations for our 
study the $J/\psi$ meson production in $\gamma \gamma$ collisions
($e^{+}e^{-}\to e^{+}e^{-} + J/\psi + X$). First of all,
recently the DELPHI collaboration has presented preliminary data
on the inclusive $J/\psi$ production in $\gamma \gamma$ collisions at
CERN LEP2~[23], which wait to be confronted with different
theoretical predictions. On the one hand it was shown recently that 
results obtained with account of the color octet contributions in the
framework of the NRQCD do not contradict the DELPHI 
data~[24]. On the other hand we know that a description of the inclusive
$J/\psi$ production in $p\bar p$ collisions at Tevatron in
the $k_T$-factorization approach needs some significantly lower
values for the color octet matrix elements to fit the Tevatron
data~[22, 25].

It is also known that the inclusive $J/\psi$ production 
in photon-photon collisions is dominated by the single-resolved 
process~[26] and therefore it reveals the gluon structure of the photon.
However the unintegrated gluon distributions in the photon 
$\Phi_{\gamma}(x,q_T^2,\mu^2)$ are poorly known 
in contrast with the similar distributions in the proton and 
no attempts have been made to describe them until recently~[27, 28]. 
Their knowledge is in particular necessary for the description
of heavy quark and quarkonium production in $\gamma \gamma$ collisions
within the semihard QCD approach. 
First application of the $k_T$-factorization approach for the case
of resolved photons in heavy quark production is performed 
in~[29] where the Kimber-Martin-Ryskin (KMR)~[30]
and Golec-Biernat-Wusthoff (GBW)~[31, 32] prescriptions for 
the unintegrated gluon distribution in the photon were used.

In the present paper we obtain in an independent way the 
unintegrated gluon distributions in the photon using the 
method proposed in Ref.~[33] by J. Bl\"umlein for the gluon distribution in 
the proton. We calculate the cross section of inclusive $J/\psi$ production 
in $\gamma \gamma$ collisions including both 
direct and resolved photon contributions and compare them 
with preliminary experimental data~[23] taken by the DELPHI collaboration
at CERN LEP2. Additionally, we give our theoretical predictions for 
$J/\psi$ polarization properties at the LEP2 conditions.

The outline of this paper is as follows. In Section 2 we present, 
in analytic form, the differential cross section for the 
inclusive $J/\psi$ production in the color singlet model with 
$k_T$-factorization and obtain the 
unintegrated gluon distributions $\Phi_{\gamma}(x,q_T^2,\mu^2)$ in the photon.
In Section 3 we present the numerical results of our 
calculations and compare them with the DELPHI data. 
Finally, in Section 4, we give some conclusions.

\bigskip
\section{Theoretical framework} \indent 

In this section we calculate total and differential cross section for 
inclusive $J/\psi$ production in $\gamma \gamma$ collisions within the 
color singlet model with $k_T$-factorization and obtain the 
unintegrated gluon distributions in the photon.

\subsection{Kinematics and cross sections} \indent

\begin{figure}[htb]
\begin{center}  
\epsfig{figure= 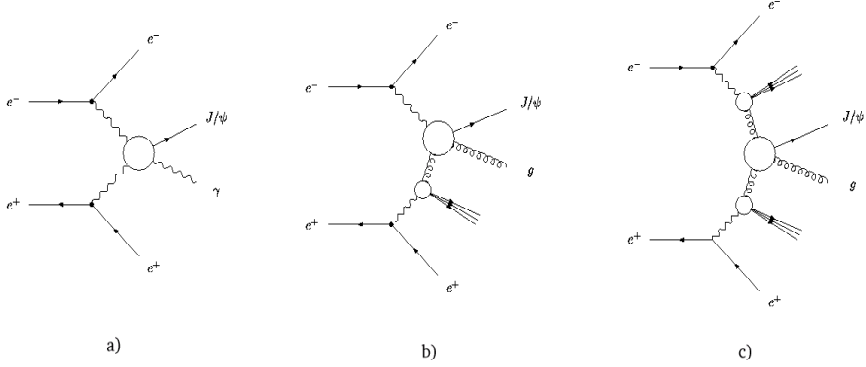,width=15.0cm,height=5.5cm}
\end{center}
\caption[]{Diagrams contributiong to inclusive $J/\psi$ production
in $\gamma \gamma$ collisions.}
\label{eps1}
\end{figure}

In $\gamma \gamma$ collisions $J/\psi$ can be produced by one of the
three mechanisms: a direct production (Fig.~1a), a photoproduction
off a resolved photon (Fig.~1b) and a by a double resolved process (Fig.~1c).
We will refer to the direct production, the once- and double resolved
photon processes as direct, 1-res and 2-res ones respectively.
The appropriate QCD motivated cross section the process
$e^{+}\,e^{-} \to e^{+}\,e^{-}\,J/\psi\,X$ is given by
\begin{equation}
\displaystyle d\sigma(e^{+}\,e^{-} \to e^{+}\,e^{-}\,J/\psi\,X) = \int f_{\gamma/e}(x_1)dx_1\, \int f_{\gamma/e}(x_2)dx_2\,d\hat \sigma(\gamma \gamma \to J/\psi\,X),
\end{equation}

\noindent where we use the Weizacker-Williams approximation for the
bremsstrahlung photon distribution from an electron~[34]:
\begin{equation}
\displaystyle f_{\gamma/e}(x) = {\alpha \over 2\pi} \left({1 + (1 - x)^2\over x}\ln{Q^2_{\rm max}\over Q^2_{\rm min}} + 2m_e^2 x \left({1\over Q^2_{\rm max}} - {1\over Q^2_{\rm min}}\right)\right)
\end{equation}

\noindent with $Q^2_{\rm min} = m_e^2 x^2/(1 - x)^2$ and $Q^2_{\rm max} = (E\theta)^2 (1 - x) + Q^2_{\rm min}$.
Here $x = E_{\gamma}/E_e$, $E = E_e = {\sqrt s}/2$, 
$\theta = 32\, {\rm mrad}$ is the angular cut that ensures the photon is 
real, and $\sqrt s = 197\,{\rm GeV}$~[23].

For partonic cross section $d\hat \sigma(\gamma \gamma \to J/\psi\,X)$ we will 
use expression obtained earlier~[20] because the 1-res processes are 
analogous to those contributing to the
lepto- or photoproduction processes at HERA, and the 2-res ones are
analogous to the $J/\psi$ hadroproduction subprocesses at Tevatron. In both cases the
unintegrated gluon distributions in the proton should be replaced by ones in the
photon.

\subsection{Unintegrated gluon distributions in the photon} \indent 

To obtain the unintegrated gluon distribution in the photon 
$\Phi_{\gamma}(x,q_T^2,\mu^2)$ we
apply the same method as for the gluon distibution in the proton
according to the prescription given in~[33]. The proposed method lies 
upon a straightforward perturbative solution of the BFKL equation 
where the collinear gluon density $xG(x,\mu^2)$ is used as the 
boundary condition. The unintegrated gluon distribution in a photon is 
calculated as a convolution of collinear gluon distribution 
$xG_{\gamma}(x,\mu^2)$ with universal weight factors:
\begin{equation}
\Phi_{\gamma}(x,q_T^2,\mu^2) = \int\limits_x^1 \,\varphi(\eta,q_T^2,\mu^2)\,{x\over \eta}\,G_{\gamma}\left({x\over \eta},\,\mu^2\right)\,d\eta,
\end{equation}

\noindent where
\begin{equation}
\displaystyle \varphi(\eta,q_T^2,\mu^2) = \cases{\displaystyle {\bar \alpha_{S}\over \eta\, q_T^2}J_0\left(2\sqrt{\mathstrut \bar\alpha_{S}\ln(1/\eta)\ln(\mu^2/q_T^2)} \right),&if $q_T^2\le \mu^2$,\cr
\displaystyle {\bar\alpha_{S}\over \eta\, q_T^2}I_0\left(2\sqrt {\mathstrut \bar \alpha_{S}\ln (1/\eta) \ln (q_T^2/\mu^2)}\right),&if $q_T^2 > \mu^2$,\cr}
\end{equation}

\noindent where $J_0$ and $I_0$ stand for Bessel function of real and 
imaginary arguments respectively, and $\bar\alpha_{S}=3\alpha_{S}/\pi$. 
In the capacity of the boundary condition we used expression for 
$xG_{\gamma}(x,\mu^2)$ from the standard GRV set~[35].
The parameter $\bar\alpha_{S}$ is connected with the Pomeron trajectory 
intercept: $\Delta = 4\bar\alpha_{S}\ln 2 = 0.53$ in the LO and 
$\Delta = 4\bar\alpha_{S}\ln 2 - N\bar \alpha_{S}^2$ in the NLO 
approximations, where  $N \sim 18$~[36]. The latter value of $\Delta$ 
have dramatic consequences for high energy phenomenology. However, some 
resummation procedures proposed in the last years lead to positive value of
$\Delta \sim 0.2-0.3$~[37, 38]. Therefore in our calculations we will
consider $\Delta$ as free parameter varying from $0.2$ to $0.53$ with
a central value at $\Delta = 0.35$.

\begin{figure}[htb]
\begin{center}  
\epsfig{figure= 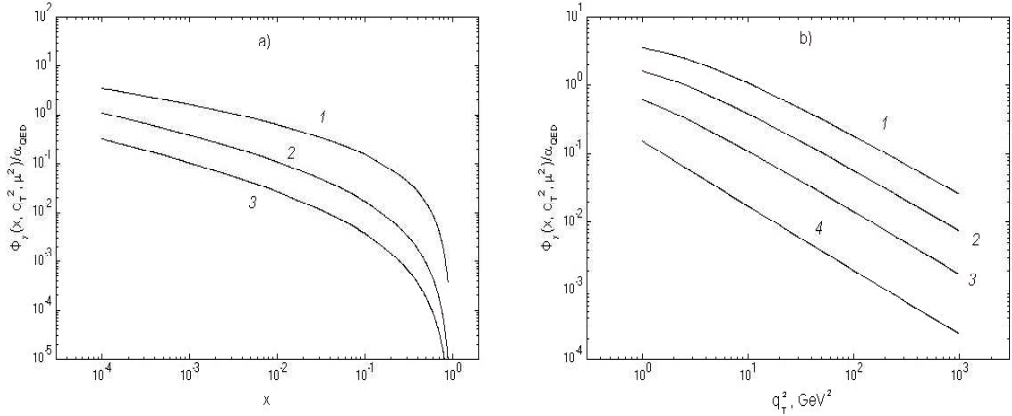,width=14.5cm,height=6.5cm}
\end{center}
\caption[]{The unintegrated gluon distribution in the photon $\Phi_{\gamma}(x,q_T^2,\mu^2)$
as a function of longitudinal momentum fraction $x$ (a) and transverse momenta 
squared $q_{T}^2$ (b) of the gluon, for fixed values of $x$ and $q_{T}^2$. 
Curves {\sl 1 --- 3} (Fig.~2a) correspond gluon virtuality 
$q_{T}^2 = 1,\,10$ and $50\,{\rm GeV}^2$ as well as curves {\sl 1 --- 4} 
(Fig.~2b) correspond longitudinal momentum 
fraction $x = 10^{-4},\,x = 10^{-3},\,x = 10^{-2}$ and $x = 10^{-1}$
 respectively. We set here $\mu^2 = q_T^2$.}
\label{eps2}
\end{figure}

We would like to note that intercept parameter $\Delta = 0.35$ was obtained from the description of $p_T$ spectrum of 
$D^*$ meson electroproduction at HERA~[39]. This value of $\Delta$ also 
was used at the analysis of experimental data on inelastic $J/\psi$ photo- and 
leptoproduction at HERA~[20], the $b\bar b$ quark production at Tevatron~[40]
and in the deep inelastic structure function $F_2^c$, $F_L^c$ and 
$F_L$ description at small $x$ region~[41, 42].

In Fig.~2 the unintegrated gluon distribution in the photon 
(3) --- (4) (so called JB parametrization) is plotted as a 
function of longitudinal momentum fraction $x$ (Fig.~2a) and 
transverse momenta squared $q_{T}^2$ (Fig.~2b) of the gluon, 
for fixed values of $x$ and $q_{T}^2$. Curves {\sl 1 --- 3} plotted
on the Fig.~2a correspond gluon virtuality $q_{T}^2 = 1,\,10$ and $50\,{\rm GeV}^2$
as well as curves {\sl 1 --- 4} plotted on the Fig.~2b correspond
longitudinal momentum fraction $x = 10^{-4},\,x = 10^{-3},\,x = 10^{-2}$ and $x = 10^{-1}$, 
respectively.

Integrating 
expression (3) --- (4) over the transverse momenta, we can obtain effective gluon distribution in
photon and compare it to experimental data~[42] taken by the H1 
collaboration at HERA. The results of such calculations are shown in Fig.~3. 
Curves {\sl 1} and {\sl 2} correspond to the gluon density in photon 
from the standard GRV set~[35] and the effective gluon density obtained 
from JB 
parametrization, respectively. One can see that both curves agree 
very well with the H1 data and at $x > 10^{-2}$ they practically coincide.

\begin{figure}[htb]
\begin{center}  
\epsfig{figure= 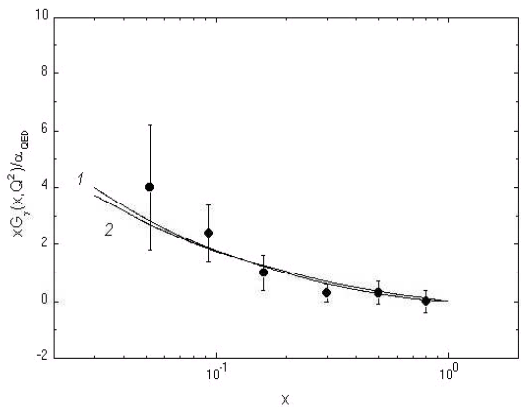,width=8.0cm,height=6.5cm}
\end{center}
\caption[]{Effective gluon distribution in
photon at $Q^2 = 74\,{\rm GeV^2}$. Curves {\sl 1} and {\sl 2} correspond 
to the gluon density 
from the standard GRV set~[35] and the effective gluon density obtained from JB 
parametrization, respectively. Experimental data are from H1~[43] 
collaboration {\LARGE $\bullet$}.}
\label{eps3}
\end{figure}

\bigskip
\section{Numerical results} \indent 

In this section we present the theoretical results in comparison with 
preliminary experimental data~[23] taken by the DELPHI collaboration 
at CERN LEP2.

There are three parameters which determine the common normalization 
factor of the cross section under consideration: $J/\psi$ meson wave 
function at the origin $\Psi(0)$, charmed quark mass $m_c$ and factorization 
scale $\mu$. The value of the $J/\psi$ meson wave function at the 
origin may be calculated in a potential model or obtained from the well 
known experimental decay width $\Gamma(J/\psi \to \mu^{+}\,\mu^{-})$. 
In our calculation we used $|\Psi(0)|^2 = 0.0876\,{\rm GeV}^3$ as in Refs.~[15, 20].

Concerning a charmed quark mass, the situation is not clear: on the one 
hand, in the nonrelativistic approximation one has 
$m_c = m_{\psi}/2 = 1.55\,{\rm GeV}$, but on the other hand there are 
many examples when smaller value of a charm mass is used in $J/\psi$ production processes at
high energies, for 
example, $m_c = 1.4\,{\rm GeV}$~[12--15, 24]. However, in our 
previous papers~[20] we analyzed in detail the influence of the
charm quark mass on the theoretical results. We found that the main
effect of a change of the charm quark mass connects with the final
phase space of the $J/\psi$ meson, and in the subprocess matrix
elements this effect is negligible. Taking into account that
the value of $m_c = 1.4\,{\rm GeV}$ corresponds to the unphysical
phase space of the $J/\psi$ state, in the present paper we will use 
the value of the charm mass $m_c = 1.55\,{\rm GeV}$ only. 

Also the most significant theoretical uncertanties come from the 
choice of the factorization scale $\mu_F$ and renormalization one $\mu_R$. 
One of them is related to the evolution of the gluon distributions 
$\Phi_{\gamma}(x,q_T^2,\mu_F^2)$, the other is responsible 
for strong coupling constant $\alpha_{S}(\mu_R^2)$. As often one done in 
literature, we set them equal $\mu_F = \mu_R = \mu$. 
In the present paper we used the following choice 
$\mu^2 = q_{T}^2$ as in Refs.~[4, 20].

\begin{figure}[htb]
\begin{center}  
\epsfig{figure= 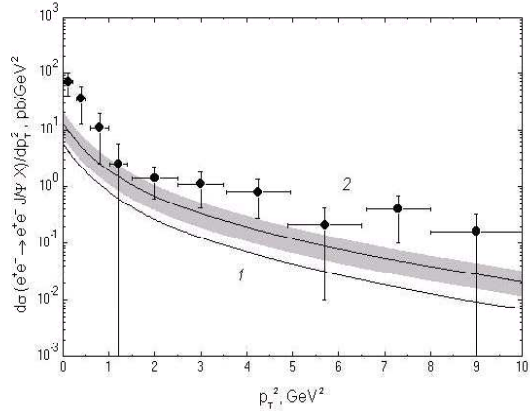,width=8.0cm,height=6.5cm}
\end{center}
\caption[]{The $J/\psi$ meson transverse momentum distribution
in comparison with DELPHI preliminary data~[23] at $\sqrt s = 197\,{\rm GeV}$
and $|y_{\psi}| < 2$. The solid lines {\sl 1} and {\sl 2}
correspond to the standard parton model (SPM) calculations at the leading 
order approximation with the GRV (LO) gluon density (at $m_c = 1.55\,{\rm GeV}$)
and the $k_T$-factorization results with the JB unintegrated gluon 
distribution (at $m_c = 1.55\,{\rm GeV}$ and $\Delta = 0.35$). 
The shaded band of the $k_T$-factorization predictions connected with the value of 
pomeron intercept $\Delta$: the lower border corresponds to $\Delta = 
0.2$ and the upper one corresponds to $\Delta = 0.53$.}
\label{eps4}
\end{figure}

The results of our calculations are shown in Fig.~4---6.
Fig.~4 displays the $J/\psi$ meson transverse momentum distribution
in comparison with DELPHI preliminary data~[22] at $\sqrt s = 197\,{\rm GeV}$
and $|y_{\psi}| < 2$. The solid lines {\sl 1} and {\sl 2}
correspond to the standard parton model (SPM) calculations at the leading 
order 
approximation with the GRV (LO) gluon density (at $m_c = 1.55\,{\rm GeV}$)
and the $k_T$-factorization results with the JB unintegrated gluon 
distribution in photon (at $m_c = 1.55\,{\rm GeV}$ and $\Delta = 0.35$). 
The shaded band is the $k_T$-factorization predictions connected with the value of 
pomeron intercept $\Delta$: the lower border corresponds to $\Delta = 
0.2$ and the upper one corresponds to $\Delta = 0.53$.
One can see that the behavior of the
transverse momentum distribution within the experimental and
theoretical uncertanties can be explained by the color singlet 
model with $k_T$-factorization\footnote{In a Ref.~[24] it was claimed
about some evidence for color octet mechanism of NRQCD from these DELPHI
experimental data~[23]. However in a recent paper~[44] the Belle collaboration
has published results on prompt $J/\psi$  and double
$c\bar c$ quark production in  $e^+e^-$-annihilation, where the additional 
$c\bar c$ pair fragments into either charmonium or charmed mesons.
It was shown that $J/\psi$ signal predicted in color octet model
is not observed and that large fraction of prompt
$J/\psi$ events is due to the $e^+e^- \to  J/\psi c\bar c$ process~[44].
We would like to note it is possible that in the DELPHI data at 
LEP2 the contributions of $J/\psi\,D^{*\,\pm}$ associated production are presented too.
The calculations of these processes are in progress.}, 
while the collinear approximation prediction
are lower than data by an order of magnitude.

\begin{figure}[htb]
\begin{center}  
\epsfig{figure= 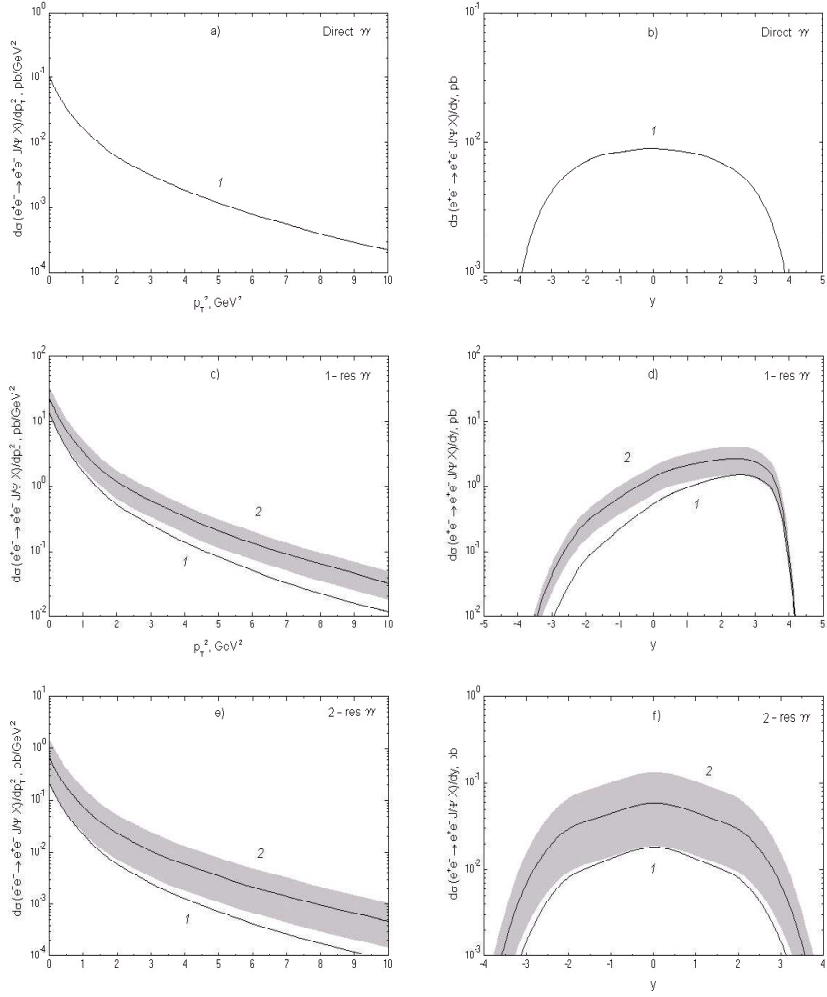,width=12.0cm,height=14.0cm}
\end{center}
\caption[]{The transverse momenta and rapidity  
distributions of the inclusive $J/\psi$ meson production 
at $\sqrt s = 197\,{\rm GeV}$, where all contribution (direct production,
1-res and 2-res processes) are plotted separately. Curves {\sl 1} and 
{\sl 2} are the same as in Fig.~4.}
\label{eps5}
\end{figure}

We would like to note a some difference in the shapes between curves 
obtained using the $k_T$-factorization approach and the colliner parton model. 
This difference manifests the $p_T$ broadening effect which is connected to
the initial gluon transverse momentum and it is usual for the $k_T$-factorization 
approach~[20, 39, 40].

Fig.~5 show the transverse momenta and rapidity  
distributions of the inclusive $J/\psi$ meson production 
at $\sqrt s = 197\,{\rm GeV}$, where all contribution (direct production,
1-res and 2-res processes) are plotted separately. Curves {\sl 1} and {\sl 2} and
shaded bands are the same as in Fig.~4. One can see that 1-res contribution 
dominates, while the direct and 2-res contributions are only a small
corrections to the 1-res one. This is contrast to the inelastic $J/\psi$ production case 
at HERA, where the resolved photon contribution is an small fraction
of the direct one. Also one can see that 2-res contributions is more 
sensitive to various  theoretical uncertainties.

As it mentioned above, the main difference between $k_T$-factorization and 
other approaches connects with polarization properties of the final 
particles. The "nonsense" polarization of the initial BFKL gluons should
result in observable spin effects of final $J/\psi$ mesons~[45].
In the present paper we calculate the $p_T^2$ dependence of the spin aligment 
parameter $\alpha$:
\begin{equation}
\alpha (p_T^2) = {d\sigma/dp_T^2 - 3\,d\sigma_L/dp_T^2\over d\sigma/dp_T^2 + d\sigma_L/dp_T^2},
\end{equation}

\noindent where $\sigma_L$ is the production cross section for the 
longitudinally polarized $J/\psi$ mesons. The parameter $\alpha$ 
controls the angular distribution for leptons in the 
decay $J/\psi \to \mu^{+}\,\mu^{-}$ (in the $J/\psi$ meson rest frame):
\begin{equation}
{d\Gamma(J/\psi \to \mu^{+}\,\mu^{-})\over d\cos\theta} \sim 1 + \alpha\,\cos^2\theta.
\end{equation}

\noindent The cases $\alpha = 1$ and $\alpha = - 1$ correspond to transverse
and longitudinal polarizations of the $J/\psi$ meson, respectively. 

\begin{figure}[htb]
\begin{center}  
\epsfig{figure= 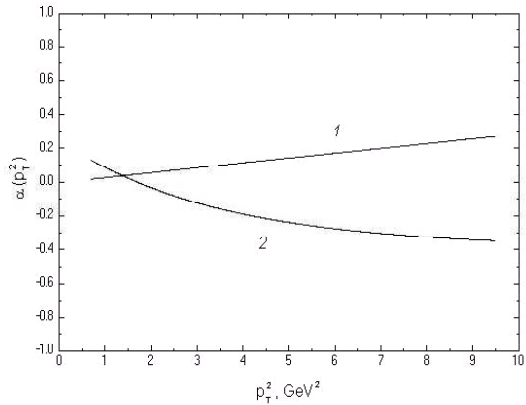,width=8.0cm,height=6.5cm}
\end{center}
\caption[]{The $p_T^2$ dependence of the spin aligment parameter $\alpha$
for the inclusive $J/\psi$ meson production at LEP2. Curves {\sl 1} and 
{\sl 2} are the same as in Fig.~4.}
\label{eps6}
\end{figure}

The results of our calculations are shown in Fig.~6. Curves {\sl 1} and 
{\sl 2} are the same as in Fig.~4. As in inelastic $J/\psi$ photoproduction 
at HERA~[20], we have large difference between predictions of the 
leading order of collinear parton model and the $k_T$-factorization 
approach. It was shown that account of the color octet contributions
do not change a predictions of $k_T$-factorization approach
for the $J/\psi$ polarization properties in $p\bar p$ and $ep$ collisions
(see~[22]).
Therefore experimental study the polarized $J/\psi$ production 
at CERN LEP2 will be an additional 
test of BFKL gluon dynamics.

\bigskip
\section{Conclusions} \indent 

In this paper we considered the inclusive $J/\psi$ meson production at 
CERN LEP2 within the colour singlet model and the $k_T$-factorization 
approach, including both direct and resolved photon contributions.
The unintegrated gluon distribution in the photon
is determined, using the Bl\"umlein's prescription for unintegrated gluon 
distribution in a proton. 
We compared the theoretical results with preliminary experimental data 
taken by the DELPHI collaboration at CERN LEP2.
We find that the $k_T$-factorization results (in contrast with the 
SPM ones) with the JB unintegrated gluon distribution in the 
photon agree with data within the experimental and
theoretical uncertanties.
Finally, it is shown that experimental study of a polarization of $J/\psi$ 
meson at LEP2 should be additional test of BFKL gluon dynamics.

\bigskip
\section{Acknowledgments} \indent 

The authors would like to thank S.~Baranov for encouraging interest and 
very useful discussions. The study was supported in part by RFBR grant 
${\rm N}^{\circ}$ 02--02--17513. A.L. also was supported by INTAS grant 
YSF'2002 ${\rm N}^{\circ}$ 399.

\end{large}
\end{document}